# EDUCATION, NEIGHBOURHOOD EFFECTS AND GROWTH:
# AN AGENT BASED MODEL APPROACH

**TANYA ARAÚJO[*] and MIGUEL ST. AUBYN**

*Research Unit on Complexity in Economics (UECE)[†]*
*ISEG - Instituto Superior de Economia e Gestão*
*Technical University of Lisbon*
*Rua Miguel Lupi 20, P-1249-078 Lisbon, Portugal*
*\*E-mail: tanya@iseg.utl.pt*

**Abstract**

Endogenous, ideas-led, growth theory and agent based modelling with neighbourhood effects literature are crossed. In an economic overlapping generations framework, it is shown how social interactions and neighbourhood effects are of vital importance in the endogenous determination of the long run number of skilled workers and therefore of the growth prospects of an economy. Neighbourhood effects interact with the initial distribution of educated agents across space and play a key role in the long run stabilisation of the number of educated individuals. Our model implies a tendency towards segregation, with a possibly positive influence on growth, if team effects operate. The long run growth rate is also shown to depend on the rate of time preference. Initial circumstances are of vital importance for long run outcomes. A poor initial education endowment will imply a long run reduced number of skilled workers and a mediocre growth rate, so there no economic convergence tendency. On the contrary, poor societies will grow less, or will even fall into a poverty trap, and will diverge continuously from richer ones.

**1. Introduction**

Recent improvements in multidisciplinary methods and, particularly, the availability of powerful computational tools, are giving researchers an increasing opportunity to investigate economies in their true complex nature. The adoption of a complex systems approach allows for modelling macroeconomic structures from a bottom-up perspective - understood as resulting from the repeated local interaction of economic agents - without disregarding the consequences of the structures themselves to individual behaviour, emergence of interaction patterns and social welfare. Agent-based models are at the leading edge of this endeavour.

Agent-based models are increasingly used in different fields of economics ([10]). Many of these models fall into the field of Finance ([11], [12], [13], [15] and [17]) and a very important part of them deals with Innovation and Diffusion processes ([5], [6], [7] and [18]). Among the later, some of the models encompass the study of economic growth. However, economic growth models do not usually account for both ideas-based macroeconomic growth and the dynamics of interactions among economic agents, in what concerns education decisions and outcomes.

Several new, endogenous growth models emphasize the role of "ideas" in driving economic growth[a]. As stated by Jones [9], ideas are nonrival goods – a good idea can be used by anyone without diminishing the possibility of its use by others. Some good ideas, or inventions, are not directly linked to economic production and growth – as is the case of the works of Shakespeare or the symphonies of

---

[*] Corresponding author.
[†] UECE is supported by FCT (Fundação para a Ciência e a Tecnologia, Portugal), under the POCTI programme, financed by ERDF and Portuguese funds.
[a] This includes the seminal works of Lucas [14] and Romer [16], and, to our view, also the Schumpeterian models where innovation occurs stochastically at a rate that depends on the quantity of labour allocated to R&D (Aghion and Howitt [1] and [2]).

Beethoven. They are, however, good examples of nonrivalry – the fact that a theatre company in New York uses the Hamlet text in a performance does not hinder any other similar event by other company in, say, Paris or Prague. Other nonrival ideas are clearly linked to economic innovation and to growth – this is the case, in the Neolithic age, of the discovery that plants could be cultivated or that animals could be domesticated, or, more recently, of electricity and the computer.

As in Jones [9], we develop a model where ideas are produced by a fraction of the working population. These ideas, or innovations, are used by the rest of the workers to produce final goods. If ideas are kept constant, than the usual replication argument is valid. That doubling other inputs leads to twice the output is suggested by the possibility of having another, identical, economy, producing exactly the same and in the same way as the one we are considering. However, when ideas are allowed to growth there will be increasing returns to scale. This derives from the nonrival nature of ideas. Doubling all factors including ideas would be the equivalent of having two more advanced economies, sharing a higher stock of concepts, and therefore producing each one more than before.

We depart from the usual solutions in this type of growth models in what concerns an important allocative social decision, concerning the share of the working population that is engaged in producing ideas. In our overlapping generations framework, ideas are invented by educated workers, which are agents that have studied in a previous period. An agent decision to study, or else, to stay uneducated, is taken following a socially conditioned economic reasoning. On one hand, each agent is concerned with his or her lifelong income, so that it may be worthwhile to give up some present income in order to become part of an education elite that is usually better paid. However, other factors may influence this decision, including a possible subjective bias towards education, and, perhaps more importantly, the so-called neighbourhood effects.

In a strict sense, there is a neighbour effect in the education decision if this one is not taken independently from the residential area where the agent lives. For example, it is sensible to assume that, *ceteris paribus*, a children's education attainment depends positively on the average human capital stock in his or her neighbourhood. In a more general sense, one can conceive a more general "social space" where a distance can be defined[b]. In this case, one can think, for example, that some norms of behaviour within a social group make some individuals more prone to follow higher education than others. In modelling terms, we place agents in a space such that a neighbourhood is defined. This neighbourhood can be interpreted in social or even familiar and/or in territorial terms. The fundamental point is that an agent's behaviour will depend on other agents close to him or her, namely on their decisions concerning education and on their outcomes[c].

It is worth noticing that some empirical studies confirm that neighbourhood effects are important in what concerns education decisions and outcomes. Durlauf [8] surveys several empirical papers where these effects are tested considering different outcomes. Education related outcomes were examined in nine of these papers, and six of them provide evidence in favour of the relevance of neighbourhood effects.

Our model therefore is a contribution to interweave two lines of research that have progressed in a separated way. These are the endogenous, ideas based macroeconomic growth models, where the representative optimizing agent is the device that allows solving different allocative decisions, and the agents based economic literature, with a strong emphasis on heterogeneity and social interactions. Introducing heterogeneous agents and neighbourhood effects into a growth model allows to identify such important phenomena as economic divergence and how some economies can even fall into a poverty trap, with income growth not occurring at all.

The rest of this paper is structured as follows. In section two we expose the main features of our education and growth model. Results from simulating the model are provided in section 3. Section 4 concludes.

---

[b] Durlauf [8] surveys the literature on the role of neighbourhoods in influencing socioeconomic outcomes. See also Akerlof [3] for more on the notion of social space.

[c] See Bénabou [4] for another modelling strategy considering neighbour effects in education.

## 2. An agent based model of education and growth – the features

### 2.1 *Population characteristics*

There are *N* individuals, or agents, in our economy. Each agent lives for two periods. Population size does not change and generations overlap. Consequently, there are at each point in time *N/2* young persons, or juniors, and *N/2* seniors.

A young person is either a student or a young worker. In the latter case, it becomes immediately part of the unskilled labour force. Otherwise, and if a young person spends the time studying, it will become a skilled worker when older. This implies that at each point in time the population is comprised of four groups:
- the young students;
- the young, or junior, unskilled workers;
- the senior unskilled workers, i. e, those that did not study in the previous period;
- the skilled workers, necessarily senior, as they have studied when young.

### 2.2 *Organization in space and decision to educate*

Agents are disposed on a ring (or one-dimensional lattice without boundary conditions) with a fixed neighbourhood size (*ng*). As such, the proximity of an individual to each of its (2*ng*) neighbours decreases smoothly in moving away either from its left or from its right. An agent does not change its location when it becomes older. However, at the end of each period, the junior agents become seniors, all senior agents die and a young agent will be born in each vacant location. Newborn agents will decide whether to start working immediately as an unskilled agent or else to become a student and to postpone its working life as a skilled worker.

The education decision is taken mixing two different perspectives. There is a neighbourhood effect – it becomes more probable that an agent becomes educated if its neighbours are skilled workers. Also, the agent takes a comparative income point of view. Studying implies a period of time without earnings, so the agent compares the present value of its income as a skilled worker with the present value of an unskilled worker income.

In formal terms, suppose that the relevant neighbourhood for each agent has size *ng*. The agent decides to study if:

$$\sum_{i=1}^{2ng} rw.DS_i > ng \qquad (1)$$

where $DS_i$ is a parameter taking value one if neighbour *i* is educated and zero otherwise; *rw* is the relative weight given to the educated. Note that, when *rw* equals one, the agent will decide in favour of education if the number of educated people in the neighbourhood exceeds the number of unskilled workers. In more general terms, though, the agent decides according to a weighted average of the number of skilled and unskilled neighbours. *rw* is a weight defined as:

$$rw(t) = \alpha \frac{ws(t-1)}{\beta(\rho).wu(t-1)}, \qquad (2)$$

where *ws(t-1)* and *wu(t-1)* denote skilled and unskilled labour wage in the previous period, respectively. $\rho$ is a discount rate, $\beta(\rho)$ a monotonous function with $\frac{\partial \beta}{\partial \rho} > 0$ and $\alpha$ an exogenous parameter[d].

The relative importance of educated neighbours increases with the skilled to unskilled wage ratio and it decreases with the discount rate. When $\rho$ increases, the future is more heavily discounted, implying that the agent is less inclined to study and to wait for future, possible higher, earnings.

---

[d] The function $\beta(\rho)$ is derived in the appendix.

**2.3 *Production***

Production of final goods depends on two inputs – the existing stock of ideas and the amount of unskilled labour. We assume a constant intratemporal labour marginal productivity, given the stock of ideas. The final goods production function is:

$$Y(t) = A(t)LU(t), \qquad (3)$$

where $Y_t$, $A_t$ and $LU_t$ denote, respectively, final goods aggregate production, the stock of ideas and the number of unskilled workers, either junior or senior. Ideas are produced by senior skilled workers. We assume that:

$$\Delta A(t) = A(t-1).\bigl(\delta LS(t) + \gamma SD(t)\bigr), \qquad (4)$$

where $LS_t$ is the number of skilled workers and $\delta$ is a parameter directly related to skilled labour marginal productivity. $SD_t$ is a distance measure defined as:

$$SD(t) = \frac{1}{LS(t)} \sum_{i,j=1}^{LS} \frac{1}{|i-j|} \qquad (5)$$

with $i \neq j$ and being thus smaller when skilled workers are located far from each other and larger in the opposite situation. The parameter $\gamma$ allows for specifying the strength of team effects.

**2.4 *Wages and income distribution***

We assume there is a social contract, so that unskilled workers get the income share that corresponds to what would have been produced if the stock of ideas remained constant. On the other hand, skilled workers get all surplus that results from new ideas. Accordingly, unskilled workers income at time $t$ is:

$$YU(t) = A(t-1)LU(t), \qquad (6)$$

Skilled workers income is equal to:

$$YS(t) = \bigl(A(t) - A(t-1)\bigr)LU(t) \qquad (7)$$

Note that these two shares add up to total income: $YS_t + YU_t = A_t LU_t = Y_t$. All income is distributed as wages. Note that there is no physical capital in the model, and therefore no physical investment. However, the decision to educate can be regarded as a choice to invest in human capital, as individuals are trading present income, that could be earned by entering the labour force without delay, by future increased income, received as skilled labour wages. The corresponding unskilled and skilled labour wages, are, respectively:

$$wu(t) = \frac{YU(t)}{LU(t)} = A(t-1), \qquad (8)$$

$$ws(t) = \frac{YS(t)}{LS(t)} = \bigl(A(t) - A(t-1)\bigr)\frac{LU(t)}{LS(t)}. \qquad (9)$$

**2.5 *Steady state, poverty trap and over education collapse***

Suppose the composition of the labour force is stable, so that the number of skilled and unskilled workers stays constant. Moreover, suppose that the disposition of skilled and unskilled agents in the ring does not change. In that case, the economy is in a steady state characterized by a constant growth rate of final goods production. To show this, note, from Eq. (4), that the steady state stock of ideas growth rate is equal to:

$$g(A_{ss}) = \frac{\Delta A_{ss}}{A_{ss}} = \delta LS_{ss} + \gamma SD_{ss}, \qquad (10)$$

where $g(.)$ denotes the growth rate and the subscript $ss$ refers to a steady state variable. $LS_{ss}$ and $SD_{ss}$ are constant by assumption, and this implies that $g(A_{ss})$ is constant.

As $LU_{ss}$ is constant, Eq. (3) implies that steady state final goods production growths at the same rate as $A$: long-term growth results necessarily from innovation (new ideas):

$$g(Y_{ss}) = g(A_{ss}) \qquad (11)$$

In addition, note from Eq. (6) that unskilled labour wages also grow at the same rate as $A$. Also, from Eq. (7), and given that the growth rate of $A$ is constant and that the ratio of skilled to unskilled labour does not change, skilled labour wages will also grow at that rate. Steady state income distribution shares are therefore constant.

Having so defined a steady state, the question of whether this economy will tend to such a long run growth path arises. If the answer is positive, one would like to know how this steady state changes if initial conditions differ. However, and as it will be shown in the next section, it is not always the case that the economy converges to a positive growth steady state. In fact, and depending on initial conditions, the economy can be driven towards two different pathological conditions, the "poverty trap" and the "over education collapse". These will be defined next.

Suppose that, for some reason, there are no skilled workers at all in the economy. In this case, no agent will ever decide to go into further education and become a skilled worker. In Eq. (1), the parameter $DS_i$ will always take the value 0, as there are no skilled senior agents in the neighbourhood of any junior agent being born. Once the number of skilled agents reaches 0, it will remain 0 forever, as no example of the success of education will push others to the same option. Consequently, $A$, the stock of ideas, will remain constant once all workers are unskilled. From Eq. (3), it is apparent that $Y$, the production of final goods, stagnates. We will call this unhappy ending as the "poverty trap".

The converse situation where all workers are skilled is perhaps more dramatic. *Mutatis mutandis*, and for the very same reasons, no unskilled worker would ever be born after the last noneducated agent died. However, one has to notice that the economy immediately collapses once there are no unskilled workers. Even if the stock of ideas has been progressing at a good rate, it is not possible to produce final goods once $LU$ is zero in Eq. (3).

**3. Simulation results**

To simulate our model, the exogenous parameters are set to the values shown in Table1, in which a baseline and six alternative scenarios are defined. In the following, besides analyzing the baseline and the different scenarios, we study with special care the possibility for an economy to fall into a **poverty trap** or else to engage in an **over education path**.

**3.1 *The baseline***

When the baseline economy starts, there are 50 unskilled agents, junior or senior. The other 50 individuals are either students, if they are junior, or skilled employees, if they are senior. In a typical, average, baseline simulation, the number of employees equals 75, 25 being skilled. The location (on the ring) of each agent, junior or senior and skilled or unskilled, is randomly determined.

Fig.1 displays the initial setting in a typical baseline simulation. Each agent is represented by a small circle: unskilled agents are coloured black, the white being students or skilled agents. As their initial location is randomly chosen, there are groups or neighbourhoods of skilled or unskilled agents of different size. An uncompanied agent is an agent whose closest neighbour in the clockwise sense has a different colour[e].

---
[e] Later in the paper we shall use the concept of uncompanied agent in order to account for segregation effects.

Table 1. Exogenous parameters.

| Parameters | Baseline | Scenarios | | | | | |
|---|---|---|---|---|---|---|---|
| | | 1 | 2 | 3 | 4 | 5 | 6 |
| Number of agents | | | | 100 | | | |
| Skilled labour productivity, $\delta$ | | | | 0.03 | | | |
| Number of periods | | | | 30 | | | |
| Relative importance of education, $\alpha$ | 1 | *2* | 1 | 1 | 1 | 1 | 1 |
| Discount rate, $\rho$ | 0.05 | 0.05 | *0.1* | 0.05 | 0.05 | 0.05 | 0.05 |
| Initial number of unskilled agents | 50 | 50 | 50 | *80* | *20* | 50 | 50 |
| Team effect in producing ideas, $\gamma$ | 0 | 0 | 0 | 0 | 0 | *0.2* | 0 |
| Neighbourhood range, $ng$ | 3 | 3 | 3 | 3 | 3 | 3 | *5* |

In the baseline, we set $\alpha=1$, so that agents are not biased towards education, in the sense that they do not give any special importance to it *de per se*. They will chose to acquire skills if and only if there are some educated neighbours that earn sufficiently more than those that remained unskilled. The skilled labour productivity parameter, $\delta$ was set to 0.03. As no team effects were considered ($\gamma=0$) and once $LS = 25$ in Eq. (1), it gives an implied initial growth rate of 75% per half generation, which seems a sensible value[f].

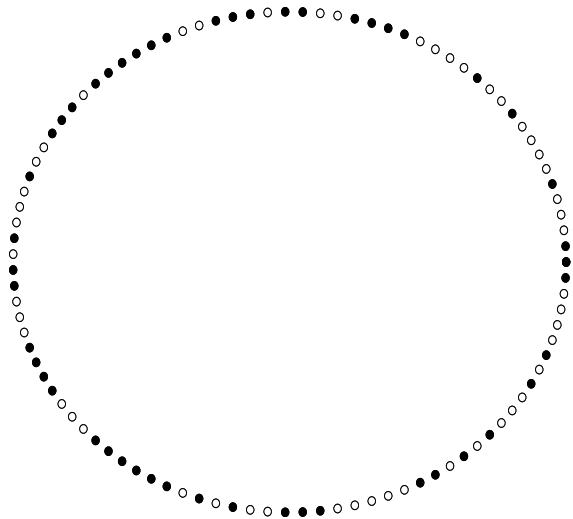

Fig. 1. Baseline, initial setting

The neighbourhood range was set to three, meaning that when an agent decides weather to pursue his or her studies by considering the status and remuneration of his or her six closest senior neighbours (three to the left and three to the right). According to Eq. (1), if there is no skilled agent in the neighbourhood, the agent will not chose to become educated, and, symmetrically, if all neighbours are educated, the agent will become a student. If the neighbourhood comprises three skilled and three unskilled agents, the education decision will be taken if the skilled labour future wage exceeds sufficiently the unskilled one so that it is worth to spend some years without a wage in present value terms. If the number of skilled agents in the neighbourhood is positive but smaller than the number of unskilled individuals, then an even greater skilled labour wage will be required in order to convince the agent to remain at school.

The baseline discount rate equals five percent, a value not very different from empirically observed real interest rates. The number of time periods for each simulation, 30, was chosen to make certain the economy reached a steady state where the number of skilled workers remains constant.

The results reported are the average results over 1000 simulations. As can be read from Table 2, the average steady state number of skilled and unskilled workers, respectively 26.0 and 48.1, is very close to initial settings. In this steady state, and from Eq. (10), the number of ideas grows at a rate equal to 0.78, as can be read from the first line.

---

[f] If half a generation lasts for, say, 25 years, this would imply a 2.3% annual growth rate.

As the number of unskilled workers is constant in the steady state, this is also the output growth rate (recall equality (11)). Also, both skilled and unskilled labour wages will grow at that very same rate. Note that the steady state skilled labour wage exceeds the unskilled wage by 44 percent, as can be read from the line "steady state relative skilled labour wage".

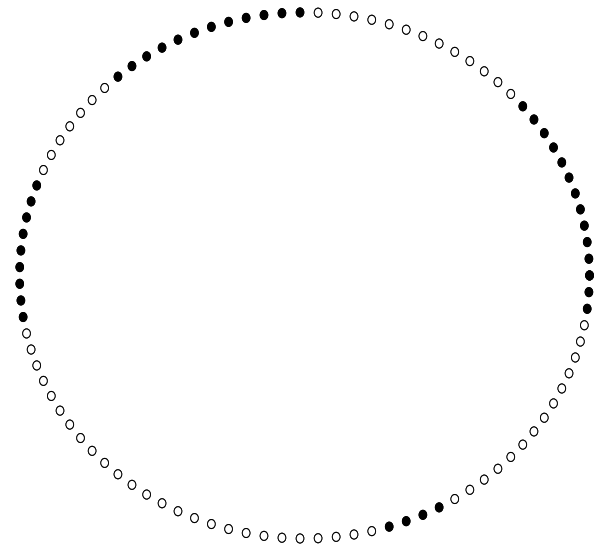
Fig. 2. Baseline, final setting

Fig. 2 displays the steady state setting of agents on the ring. As time goes by, there is a segregation mechanism going on. Agents that are born close to skilled individuals tend to go into further education and persons close to unskilled agents have a higher probability of not pursuing their studies. The long run effect is the one described in Fig.2: skilled agents are clustered in a small number of neighbourhoods, a very different pattern when compared to the random initial distribution depicted in Fig.1.

It is possible to summarize the degree of clustering by counting the number of observed partitions, which corresponds to the number of uncompanied agents existing in the ring. The initial and steady state number of partitions are presented in Table 2. The initial average number of partitions is 50, which decreases drastically to six.

**3.2 *Scenario 1 – Education is important***

In this variant the parameter $\alpha$ takes the value two in Eq. (2). This implies that, *ceteris paribus*, the weight put on education doubles. In more precise terms, this means that one skilled wage unit is valued as two unskilled wage units. Agents value education *de per se*, in the sense that they dislike less the time working in an activity that requires their acquired skills. This change in preferences implies that the economy settles in a steady state that is significantly different from the baseline. The number of skilled workers is higher (39.0), and, consequently, the steady state output growth rate is also higher (1.17).

However, and due to the relative abundance of educated agents, the skilled labour remuneration is now smaller than the unqualified labour wage (in Table 2, the long run ratio of skilled to unskilled labour wage equals 0.665). Unskilled labour earns a premium, as it is a relatively undesirable activity. Again, and even more than in the baseline, educated agents tend to cluster in three or four neighbourhoods – the average final number of partitions equals 3.33.

Table 2. Simulation results. Average results over 1000 simulations.

|  | Baseline | Scenarios | | | | | |
|---|---|---|---|---|---|---|---|
|  |  | 1 | 2 | 3 | 4 | 5 | 6 |
| Steady state Y growth rate | 0.78 | 1.17 | 0.62 | 0.39 | 0.94 | 1.98 | 0.76 |
| Steady state number of unskilled employees | 48.1 | 22.1 | 58.3 | 73.7 | 37.4 | 29.4 | 49.5 |
| Steady state number of skilled employees | 26.0 | 39.0 | 20.8 | 13.0 | 31.4 | 35.3 | 25.2 |
| Steady state relative skilled labour wage | 1.44 | 0.665 | 1.75 | 2.21 | 1.122 | 1.64 | 1.48 |
| Initial number of partitions | 50.5 | 50.6 | 50.6 | 32.2 | 32.1 | 50.5 | 50.4 |
| Steady state number of partitions | 6.00 | 3.33 | 4.35 | 5.5 | 3.00 | 5.35 | 3.47 |

### 3.3 *Scenario 2 – Future is less valued*

The discount rate increases from five to ten percent in this scenario. Future wages are more heavily discounted. Consequently, investing in education becomes less attractive, as one of its motivations are the higher future earnings that compensate present thrift, as students abstain from wage income when studying. Comparing to the baseline, a smaller number of agents will chose education, and the steady state number of skilled workers is lower (20.8, compared to 26.0 in the baseline). This affects the long run output growth rate, which declines from 0.78 to 0.62. As in the baseline, skilled workers cluster in a limited number of neighbourhoods, the partition number being similar. Finally, it is worth telling that the relative wage of skilled workers is now higher, a logical consequence of its relative scarcity.

### 3.4 *Scenario 3 – The initial number of unskilled agents is higher*

Initial conditions matter. As it will be shown later, when the initial number of educated people is too small, the probability of extinction becomes non-negligible and the economy may fall into stagnation. For an initial number of unskilled workers of 80, as imposed in this scenario, this probability is negligible. However, this economy, that starts uneducated, tends to stick into a mediocre growth path. The steady state number of uneducated workers averages 73.7, much higher than the baseline value (48.1). Note that the corresponding growth rate is smaller (0.39), and that skilled agents wage exceeds the double of the unskilled labour one. In this relatively undeveloped economy, skilled agents are a small elite. Even if they are well paid, a relatively small number of young people enters this small social group that does not show up in spatial terms.

### 3.5 *Scenario 4 – The initial number of unskilled agents is lower*

Scenario 4 is the converse of scenario 3. Here, the initial number of unskilled agents, 30, is lower than in the baseline. The economy starts as a more educated one. The fact that skilled people predominate imply that more young people tend to imitate their social environment, resulting in a steady state where the number of skilled employees, 31.4, is not very different from the number of unqualified ones, 37.4. The fact that skilled workers abound implies that their steady state wage approaches the unskilled labour one, the relative wage being now 1.122. As opposed to scenario 4, this is a relatively developed economy where education is widespread, social differences are less important, new ideas are regularly produced and the growth rate, 0.94, is stronger. As it was mentioned before and will be shown later in more detail, a high initial number of skilled workers may imply that a steady state is not attained, the economy reaching a point where unskilled agents extinguish and the economy collapses. This happened in 13 percent of computer-generated cases in this scenario. Computed averages do not take in these collapse cases.

### 3.6 *Scenario 5 – Team effect in producing ideas*

The baseline does not incorporate team effects – the same skilled workers produce the same number of ideas, irrespective of their dispersion. In scenario five, we include a team effect, such that production of ideas increases when skilled workers are close to each other. In terms of Eq. (4), we make $\gamma = 0.2$ so that when $SD(t)$ is higher, skilled agents work together, share their thoughts and give rise to more new ideas.

Comparing to the baseline, increased skilled labour productivity gives rise to higher relative skilled labour wages (1.64 comparing to 1.44). Higher wages induce more education choices from agents, and the steady state average number of skilled agents, 35.3, is therefore higher. As in scenario one or four, more skilled agents lead to an increased steady state rate of growth. Here, this outcome is reinforced by the prevalence of team effects.

### 3.7 Scenario 6 – The neighbourhood range increases

In this scenario the neighbourhood range increases from three to five, i. e., *ng* takes the value five in Eq. (1). When taking an education decision, the agent increases his or her observation range, and looks into the five, and not three, neighbours to the left and to the right. Results are very close to the baseline, but for one outcome – segregation is stronger, the average number of partitions becoming inferior (3.47, compared to five).

### 3.8 *Initial number of unskilled workers, poverty trap and over education*

Table 3 contains average results over 1000 simulations for different initial conditions in what concerns the number of unskilled agents. All other parameters are equal to baseline values. The number of initial unskilled agents was made equal to 10, than to 15, etc., till 90, and the percentage of stagnation cases and collapses was retained. Moreover, in the remaining cases of convergence to the steady state, the long run number of skilled workers and the implied growth rate are also presented.

Table 3. Initial number of unskilled workers, poverty trap and over education.

| Initial number of unskilled workers | Percentage of stagnation events (poverty trap) | Percentage of collapses (over education) | Average number of steady state skilled workers (excluding stagnation and collapses) | Implied growth rate |
|---|---|---|---|---|
| 10 | 0 | 54.2 | 33.9 | 1.017 |
| 15 | 0 | 30.2 | 32.6 | 0.978 |
| 20 | 0 | 13.0 | 31.6 | 0.948 |
| 25 | 0 | 5.2 | 30.5 | 0.915 |
| 30 | 0 | 2.2 | 29.3 | 0.879 |
| 35 | 0 | 0.7 | 28.3 | 0.849 |
| 40 | 0 | 0.1 | 27.9 | 0.837 |
| 45 | 0 | 0 | 27 | 0.810 |
| 50 | 0 | 0 | 26 | 0.780 |
| 55 | 0 | 0 | 24.7 | 0.741 |
| 60 | 0 | 0 | 23.3 | 0.699 |
| 65 | 0 | 0 | 21.9 | 0.657 |
| 70 | 0 | 0 | 20 | 0.600 |
| 75 | 1.1 | 0 | 16.9 | 0.507 |
| 80 | 1 | 0 | 13 | 0.390 |
| 85 | 5.5 | 0 | 9.6 | 0.288 |
| 90 | 20.2 | 0 | 6.9 | 0.207 |

Fig.3 summarises the percentage of stagnation and collapse events as a function of the initial number of unskilled workers.

When the economy starts highly educated, i. e., when the number of unskilled agents is low, the collapse probability may be quite high. As a great majority of the working population is skilled, it becomes possible that social imitation mechanisms are so strong that all population becomes skilled in finite time, and no one actually produces final goods. This probability of an over education trap becomes unimportant, or smaller than five percent, for a value of initial unskilled workers higher than about 25.

Conversely, when the population is essentially unskilled, there is a considerable chance for skilled workers to extinguish, as they do not show up sufficiently in any neighbourhood, and therefore do not reproduce. Recall that when skilled workers stop existing, new ideas are not produced anymore and the economy ceases to growing. From table 3, it is apparent that the probability of stagnation becomes significant, or higher than five per cent, when initial values for unskilled workers exceed about 85.

The two last columns of Table 3 contain the average steady state number of skilled workers excluding collapses or stagnation events, and the implied growth rate, as given by Eq. (10). The number of steady state skilled workers, and therefore the long run growth rate, depends strongly on initial conditions. An economy that starts poor, with an unskilled population, will not converge to the growth path of an economy that starts with an educated population. The long run *growth rate*, and not only income level, depends negatively on the initial number of unskilled workers.

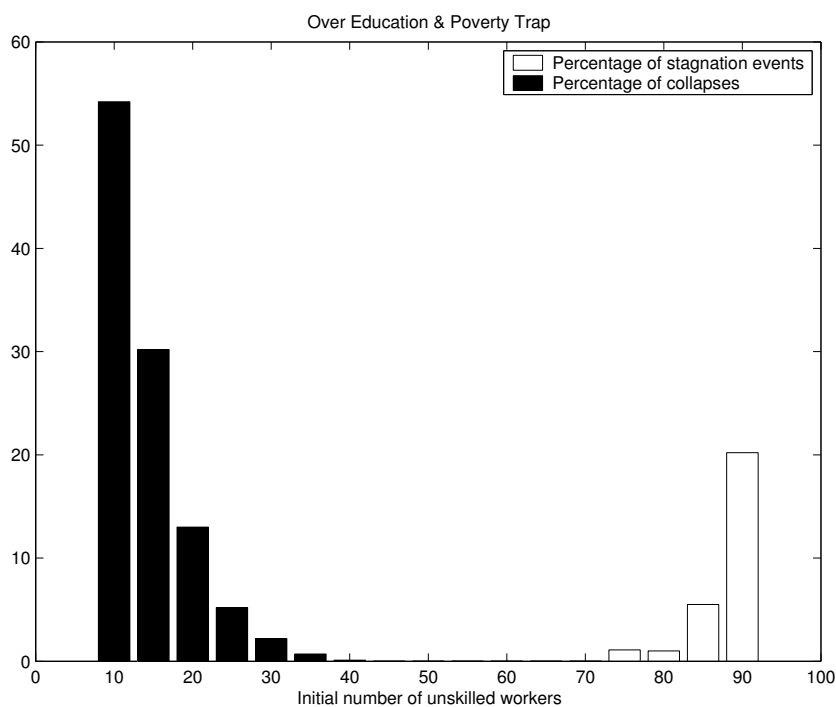

Fig. 3. Over Education and Poverty Trap.

**4. Conclusions**

This paper has shown that is it possible and fruitful to cross distinct lines of research – endogenous, ideas-led, growth theory and agent based modelling with neighbourhood effects. On the one hand, it is appealing and convincing to consider that growth, in the long run, depends essentially on innovation, and that innovation comes from new ideas, produced by educated agents. On the other hand, several empirical studies have shown that people tend to become more educated if they are born in a favourable environment, where most people are already educated, and where education is a profitable investment, in the sense that it entitles individuals to a higher income in the future.

Our economic model takes these different hypothesis seriously and in a complementary way. In an overlapping generations framework, we have shown how social interactions are of paramount importance in the endogenous determination of the long run number of skilled workers and therefore of the growth prospects of an economy. Neighbourhood effects interact with the initial distribution of educated agents across space and play a key role in the long run stabilisation of the number of educated individuals. Our model implies a tendency towards segregation – similar agents tend to live close to each other. This segregation may have a positive influence on growth, if team effects operate and ideas are best generated when people have close contact. Moreover, and as in other growth endogenous growth models, the long run growth rate depends on parameters describing agents' preferences. Namely, if they heavily discount future income or do not love education, the steady state growth rate tends to be lower.

We have also shown how initial circumstances are of vital importance for long run outcomes. A poor education endowment may mean that skills are not reproduced, all workers become unskilled, no new ideas are produced, and the economy stagnates in the long run. Even when stagnation does not occur, it is usually the case that the long run outcome will be characterized by a socially divided society, where a small number of relatively well paid skilled workers only assure a mediocre growth rate. If starting conditions are such that skilled workers already abound, the long run will be such that skilled workers will be much less scarce, their wage will be relatively lower in comparison to unskilled labour, and the stock of ideas and final goods production will grow at a higher pace. Therefore, our model does not display any kind of economic convergence tendency. On the contrary, poor societies will grow less and will diverge continuously from richer ones.

**Appendix**

At the beginning of period *t*, an agent has perfect knowledge of period *t-1* wages, namely *ws(t-1)* and *wu(t-1)*, the skilled and unskilled labour wage, respectively. Assume that agents take these values as the ones that will prevail in the future, and, for the sake of simplicity, denote them by *ws* and *wu*.

Suppose skilled agents spend nine more years at school than unskilled agents. For example, one can think they spend two more years at secondary school, four additional years to take a first degree, and finally three more years in some form of post-graduate studies. *PVS*, the present value of future wages for an agent that is starting his or her education to become skilled is then:

$$PVE = ws\left[(1+\rho)^{-9} + (1+\rho)^{-10} + \ldots + (1+\rho)^{-\tau}\right], \tag{A.1}$$

where $\rho$ is a rate of time preference or discount rate, and $\tau$ is the number of years to the end of active life, likely to be comprised between 45 and 50.

At the same time a future skilled agent starts his or her education, unskilled agents start working. With the hypothesis above, this means they work nine more years when compared to skilled workers. Let *PVU* be the present value of all wages earned by unskilled workers:

$$PVU = wu\left[1 + (1+\rho)^{-1} + \ldots + (1+\rho)^{-\tau}\right], \tag{A.2}$$

Comparing Eq.s (A1) and (A2), it is apparent that *ws* must be greater than *wu* for there to be any chance of *PVE* being greater than *PVU*. In this case, and from a pure income perspective, i. e., taking aside any subjective preference for education, the agent would chose to proceed into further education and not to remain unskilled. Let $\beta$ be the ratio between *ws* and *wu* that makes the present value of skilled labour wages equal to the present value of unskilled labour wages:

$$\frac{ws}{wu} = \beta \Rightarrow PVE = PVU \tag{A.3}$$

From Eq.s (A1), (A2) and (A3), it gives:

$$\beta = \frac{1 + (1+\rho)^{-1} + \ldots + (1+\rho)^{-\tau}}{(1+\rho)^{-9} + \ldots + (1+\rho)^{-\tau}} = \frac{A}{B}, \tag{A.4}$$

with $A = 1 + (1+\rho)^{-1} + \ldots + (1+\rho)^{-\tau}$ and $B = (1+\rho)^{-9} + \ldots + (1+\rho)^{-\tau}$. Is is easy to show that $A = \frac{1 + \rho - (1+\rho)^{-\tau}}{\rho}$ and that $B = \frac{(1+\rho)^{-8} - (1+\rho)^{-\tau}}{\rho}$. Replacing *A* and *B* in expression (A4) and simplifying, it gives:

$$\beta = \frac{(1+\rho)^9 - (1+\rho)^{8-\tau}}{1 - (1+\rho)^{8-\tau}}. \tag{A5}$$

Note that $\beta$ approaches $(1+\rho)^9$ when $\tau$ tends to infinity, and that $\beta$ is an increasing function of $\rho$. When the discount rate is higher, the future is less valued, and therefore the skilled labour wage has to be higher for agents to become indifferent between acquiring skills through education and to remain unskilled. In our simulations, we set $\tau = 48$.